# Coherency Detection and Network Partitioning based on Hierarchical DBSCAN

Faycal Znidi
Electrical and Computer
Engineering Department
Texas A&M University
Texarkana, TX, USA
fznidi@tamut.edu

Hamzeh Davarikia
Electrical and Computer
Engineering Department
McNeese State University
Lake Charles, LA, USA
hdavarikia@mcneese.edu

Mohammad Arani
Systems Engineering Department
University of Arkansas
at Little Rock
Little Rock, AR, USA
mxarani@ualr.edu

Masoud Barati
Electrical and Computer
Engineering Department
University of Pittsburgh
Pittsburgh, PA, USA
masoud.barati@pitt.edu

*Abstract*— **After a sudden disturbance, the energy balance of generators is disturbed, and the power outputs of synchronous generators vary as their rotor angles shift from their equilibrium points. This trend essentially presents the versatile response of each machine to the disturbance. Because of this change, the phase angle of the bus also differs. Hence, the versatile response of each machine can be assessed by the phase angles change at the buses close to the synchronous generator. This paper introduces a new methodology for discovering the degree of coherency among buses using the correlation index of the voltage angle between each pair of buses and use the Hierarchical Density-Based Spatial Clustering of Applications with Noise (HDBSCAN) to partition the network into islands. The proposed approach also provides the network integrity indices (connectivity, splitting, and separation) for studying the dynamic nature of the power network system. The approach is assessed on an IEEE-39 test system with a fully dynamic model. The simulation results presented in this paper demonstrate the efficiency of the proposed approach.**

## I. Introduction

The power network is routinely subjected to a range of disruptions and may become quickly unstable which my lead to a catastrophic blackout because of cascading failure. Controlled islanding can be used as a corrective measure to avoid undesirable adverse implications. Literature reviews aim to tackle this problem by using the minimal load shedding as the primary objective function. However, to find a desirable cutset the solution should satisfy a certain set of features i.e., coherent groups of generators, the minimum power flow disruption, thermal limits of transmission lines, voltage limits, transient stability, etc. [1]. The aforementioned features can be coordinated along with additional corrective measures to obtain applicable islands in the power system that will help to reduce the complexity of the islanding solution, which is similar to the kind of 0-1 knapsack problem [2]. Generally, the process of choosing the best alternative for reaching the objective is not always a simple procedure. The ideal approach would involve training all feasible combinations of the various attributes and subsequently attempt to rank all the possible combinations to find the optimal solution. Though the number of features or dimensions grows exponentially as the number of attributes increases, thus the problem can very quickly become intractable in the so-called NP-hard problems. Instead, we may select the most relevant attributes, which have the greatest impact on the islanding boundaries' formation. This set of selected attributes, in coordination with other corrective actions, can be used to create applicable islands in the power system that will help to reduce the complexity of the ICI, which is similar to the kind of 0-1 knapsack problem [2].

Many combinatorial optimization methods have been introduced in the last decade to solve the islanding problem. To search for appropriate islands multi-objective optimization techniques have been utilized in [1], where the minimal power imbalances are used as the objective function and the generator coherency as the sole constraint. In [2], the authors used the Mixed Integer Linear Programming (MILP) algorithm to obtain feasible islanding solutions. In [3], the authors introduced a two-phase methodology built on Ordered Binary Decision Diagrams (OBDD) to find an islanding solution. In this method, the islands were found where strategies must satisfy the power balance state. In [4], by utilizing the graph theory techniques the splitting strategies problem was transformed into a graph partition problem. Then, Mixed-Integer Linear Programming (MILP) was utilized to discover the islanding solution. In [5], based on dynamic frequency variations signals of the buses caused by sudden supply/demand variations, a dynamical approach was presented to discover the coherent groups of generators and the splitting strategies. One of the greatest benefits of this methodology of finding islanding solutions is that the coherency of generators was determined using the online approach and using the center of the inertia approach which has shown its benefits in numerous applications.

Spectral clustering approaches have been used in the works of literature to determine a possible islanding solution to prevent the beginning of wide-area blackouts. In [6- 8], the islanding solution strategies were determined using the constrained spectral k-embedded clustering algorithm, where the least power flow disruption represents the objective function and considering the generator's coherency as the sole constrained. In [9] an agglomerative hierarchical clustering technique is utilized to identify the coherent generators. Employing this technique to area recognition problem requires modification because it is essential to consider the interconnection of buses for evaluating the variations.

The islanding approach in the AC-DC microgrid and distribution system is proposed in [10-12], where the author pursues control actions. While the islanding can be utilized as a corrective measure, it can be used as a type of intrusion to the grid, and advanced data security approaches [13-16], as well as problem solving approaches in the collaborative environment [17-19], should be pursued to avoid potential damages to the society.

This paper introduces a new methodology for identifying the coherency between buses using the correlation index of the voltage angle at the bus to measure the strength of the association between each pair of buses. In this methodology, the versatile response of distributed generators is assessed by the phase angles variation at the buses nearby the generator. The Hierarchical Density-Based Spatial (HDBSCAN) clustering is used to partition the network into islands. The HDBSCAN does not require one to specify the number of clusters in the data a priori, as opposed to k-means. Further, the power network indices are discovered, and the strength of the buses coherency, power network integrity, and a complete system separation condition are examined.

## II. COHERENCY DECTTION BASE ON PEARSON PRODUCT-MOMENT CORRELATION COEFFICIENT

After abrupt power losses on the power grid, the energy balance of the synchronous machines is disturbed, and the rotor angular differences undergo fast changes. Then, the equilibrium becomes unstable, which essentially represents the dynamic behavior associated with every synchronous generator to the disruption. Because of this change, the phase angle of the bus also differs. Consequently, by using the phase angles dissimilarity at the buses nearby the generator, then the dynamic response of distributed the synchronous generators can be assessed. Therefore, by measuring the variation among the phase angle, by letting $i$ and $j$ as the phase angle difference between the $i^{th}$ phase angle and the $j^{th}$ phase angle, the coherency among buses can be assessed by measuring the variations among the phase angle dissimilarity [20,21]. The variation among the phase angle can be described as in (1).

$$s_{i,j} = \int_0^T \left(\Delta\theta_i(t) - \Delta\theta_j(t)\right) dt \quad (1)$$

where $s_{i,j}$ is the variation index among bus $i$ and $j$, $T$ is the examination time, and $\theta_i$ and $\theta_j$ represent the phase angles, respectively, at busses $i$ and $j$. The total electrical energy delivered or required over a given time is reflected directly by the rate of bus phase angles change at the buses [11-16]. The machine rotor speed is tightly linked to the frequency throughout the system. These frequencies correspond to the dynamic response of each synchronous machine following a major disruption. Further, these particular frequencies elements can be obtained utilizing Discrete Fourier Transform (DFT) as shown in equation (2) below:

$$F_i(f) = \sum_{k=0}^{N-1} \omega_i(k) e^{-\left(j\frac{2\pi f k}{N}\right)} \quad f = 0,1,\dots,N-1 \quad (2)$$

$$\omega_i(k) = \frac{\theta_i(k) - \theta_i(k-1)}{\Delta t} \quad (3)$$

where $F_i(f)$ symbolizes the Fourier Transform (FT) of a particular angular frequency, $\Delta t$ is the time difference among two successive samples, and $\omega_i(k)$ signifies the angular velocity of the synchronous generator $i$ at instantaneous time $k$. The time difference should be held stead during simulations. Therefore, The $N_B \times N$ matrix $F$ can be defined as:

$$F = \left[F_1(f), \dots F_i(f), \dots F_{N_B}(f)\right]^T \quad (4)$$

where $F_i$ symbolize the vector space and $N_B$ is the over-all number of busses of the electric network. Phasor measurement units (PMUs) can be used to test the angular velocities at greater rates to obtain extra frequencies samples.

Pearson Product-moment Correlation Coefficient (PPCC) is the most widely used correlation to evaluate the degree of the relationship among linearly associated variables. Normally, the correlation coefficient (CC) varies between -1.0 and +1.0 and measures the intensity and direction of the linear relationship that exists among the two quantities. About the power network, the variables represent the coherency of two different buses where a value of +1.0 indicates there is an ideal positive association among the two buses [28-30]. In order, to evaluate the degree of coherency between two different buses, the PPCC among bus $i$ and $j$ is formed as follows:

$$PPCC_{i,j} = \frac{\sum_{f=1}^n \left[\left(\theta_i - \theta_i^{avg}\right)\left(\theta_j - \theta_j^{avg}\right)\right]}{\sqrt{\sum_{f=1}^n (\theta_i - \theta_i^{avg})^2 \sum_{f=1}^n \left(\theta_j - \theta_j^{avg}\right)^2}} \quad (5)$$

In equation (5), the $PPCC_{i,j}$ is the CC between buses i and j, $n$ represent the quantity of each element of frequency and, $\theta_i^{avg}$ represents the total average of voltage angle between buses. The PPCC among two buses is a complex number because the component of $F$ is complex quantities. Though, any change in phase angle will have a substantial impact on the PPCC, while the change in the strength of two varying signals or the DC offset does not influence the PPCC. Therefore, the absolute value of complex quantities of the PPCC among two coherent buses must be 1 and 0 or ($1\angle 0$) correspondingly. Figure (1) shows the correlation coefficient in terms of phase change.

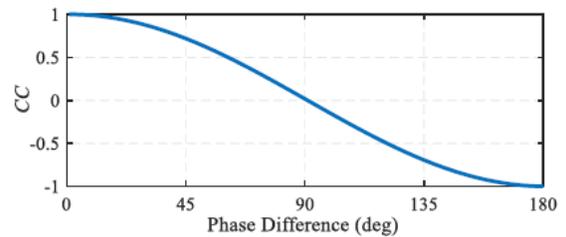

Fig. 1. Correlation Coefficient (CC) in terms of phase change

Therefore, the coherency between two buses can be revealed by the measurement of the correlation of the phase angle variation of buses. While PPCC shows the strongest relationship among two buses, the greater coefficients illustrate stronger connections indicating higher coherency between each pair of buses. As we are only looking for the perfectly correlated buses to form the PPCC likeness Matrix (PPCCsm), we only consider non-negative parameters and make others equal to zero. The PPCCsm is defined in (6 and 7).

$$\text{PPCCsm} = \begin{pmatrix} c_{1,1} & \cdots & c_{1,N_B} \\ \vdots & \ddots & \vdots \\ c_{N_B,1} & \cdots & c_{N_B,N_B} \end{pmatrix} \quad (6)$$

$$\text{PPCCsm}_{ij} = \begin{cases} CC_{i,j} & CC_{i,j} \geq 0 \\ 0 & \text{otherwise} \end{cases} \quad i,j \in N_B \quad (7)$$

For coherency identification, the buses are considered as the vertices and the PPCC of the frequency components among the buses serve as weights of the connections among those vertices. Therefore, using the edge-weighted graph, where the similarity among buses represents the edge weights, the challenge of discovering the weakest dynamic connectivity coupling among the buses can be transformed into a similarity graph-cut problem. HDBSCAN clustering techniques are employed to solve the graph-cut problems.

### III. DENSITY-BASED SPATIAL CLUSTERING APPROACH NETWORK PARTITIONING

Using the tight coherency concept discussed in section II, the phase angle among the buses in the same area would have a reasonably identical difference. Therefore, the maximum degree of coherency occurs when the PPCC among two pair buses is 1∠0 and this could be evaluated by determining the PPCC among every pair of buses within the area utilizing equation (5). The degree of association between two buses is maximal if they belong to the same group (cluster) and minimal otherwise. Buses that belong to the same group have a high level of correlation or at least higher than what they have with the rest of the network. Clustering is a machine learning practice that aims at grouping such that buses within a group are similar to each other and dissimilar from the buses in other groups. The HDBSCAN has been widely utilized for partitioning the dataset and is employed in this work to solve the clustering problem [31]. DBSCAN is a recent algorithm developed by some of the same people who write the original DBSCAN paper. The HDBSCAN is a recent algorithm formed by the same authors who developed the original DBSCAN [32]. DBSCAN works very well for clustering and can detect clusters with different densities. Generally, the complexity of DBSCAN is $O(n^2)$ in the worst case, in many practical cases much faster and it practically becomes more severe in a higher dimension

The objective of HDBSCAN is to separate all points in a data set $\boldsymbol{X} = \{\boldsymbol{x_1}, \dots, \boldsymbol{x_N}\}$ containing $n$ data objects distributed in the space according to the density/sparsity. Moreover, let D be an $n \times n$ symmetric matrix including the distances $d(\boldsymbol{x_p}, \boldsymbol{x_q})$ among pairs of objects of $\boldsymbol{X}$. The basic idea of HDBSCAN after clustering, each data point in the set either develops to a core point fitting into a cluster or it develops to a noise point as follows:

**Core points**: A point $x_p$ is a core point w.r.t $\epsilon$ and $m_{pts}$ if $|N_\varepsilon(x_p)| \geq m_{pts}$, where $N_\varepsilon(x_p) = \{x \in X/d(x, x_p) \leq \varepsilon\}$ and $|.|$ symbolize cardinality. A point in the data set will be labeled as noise if it is not a core point.
**ε-Reachable**: Two core points $x_p$ and $x_q$ are ε-reachable w.r.t $\varepsilon$ and $m_{pts}$ if $x_p \in N_\varepsilon(x_q)$ and $x_q \in N_\varepsilon(x_p)$.
**Cluster**: A cluster C w.r.t $\varepsilon$ and $m_{pts}$ is a non-empty maximal subset of $\boldsymbol{X}$ so that each pair of elements in C is density-linked.

The HDBSCAN converts the space according to the density/sparsity, from the distance weighted graph builds the minimum spanning tree by linking all data elements, and condenses the cluster hierarchy based on minimum cluster volume. Therefore, the HDBSCAN starts by decreasing $\varepsilon$ value and at the same time, it removes all edges from $m_{pts}$ having weights greater than $\varepsilon$. Then, a hierarchical cluster tree is developed up until every one of the data points in the data set is categorized as noise. Therefore, HDBSCAN can identify clusters of various densities. An HDBSCAN illustration is showing in Figure (2) below.

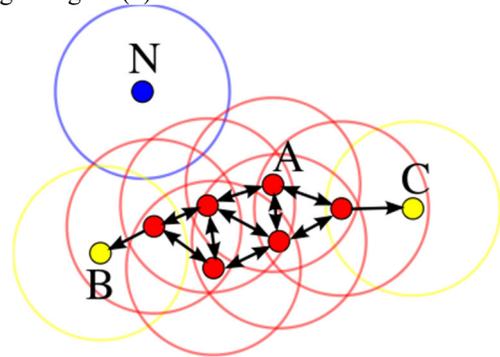

Fig. 2. HDBSCAN illustration

In Figure (2), $m_{pts}$ = 4. As can be seen from Figure (2) [33], all the redpoint are core points because the region near these points in a ε radius contains a minimum of 4 points. Further, these points form a similar cluster since they are all reachable from each other. Points B and C are not core points however, they are within reach from A i.e., through other core points, therefore these points fit into the same cluster as A. Point N is considered as a noise point since it is not directly reachable or a core point.

### IV. SIMULATION TEST CASES

To show the performance of our methodology two scenarios were created, stable case and unstable case. Furthermore, the recommended scheme estimation indicators are computed, and their trend throughout the simulation examinations is investigated. The number of generators can be accomplished by employing the spectral clustering to the on the PPCC matrix. The power grid network integrity indices could be achieved by a carry-out additional assessment on the grouped PPCC Matrix. Group Coherency Index (GCI) variation, is called as the mean of diagonal of the PPCC matrix, which presents coherently strong generators within the groups. The Group Separation Index (GSI) is defined as the mean of PPCC matrix off-diagonal which unveils to the extent that the generators in different groups tend to swing against the other groups after a disturbance.

The methodology efficacy is assessed via the simulation study performed on the modified IEEE 39-bus system shown in Figure (3). The approach has been executed in MATLAB and all time-domain simulations are attained in DIgSILENT PowerFactory. Table 1 list the events that occurred as a result.

Table 1. Events Occurred in The IEEE 39-Bus System

| Time (sec) | Description |
|---|---|
| 3.00 | Short circuit on lines 3-4 |
| 10.00 | Short circuit on lines 13-14 and 16-17 |
| 15.20 | Switch event |

Three short circuits (SC) events occurred in lines 3-4 at t=3 sec and lines 13-14 and 16-17 at t= 10 sec. Figure (4), and (5) demonstrates the generator's rotor angle and the system frequency correspondingly, which indicates the system instability after the second event.

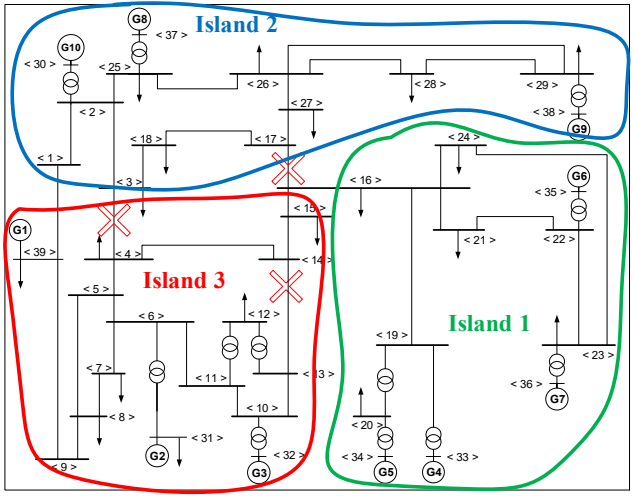

Fig. 3. EEE 39 bus system. The crosses are the event in the sample case study

Figure (6) and (7) show the variation of the GCI and GSI respectively following the events. As can be seen, after the first event at t=3s the GCI and GSI are slightly increased because the events cause the generators to swing, however, this failure caused total separation at 25.25 sec. The GCI and GSI have small variations because the generators stabilized between t=3 sec and t=25 sec. After the second event at t=20.25 sec, the GCI and GSI become unstable.

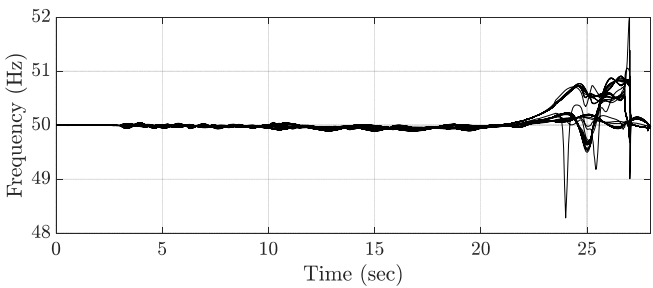

Fig. 4. The bus frequencies

The recommended solution methodology is employed to the scheme, to establish the islanding borders. Then, by evaluating the dynamic performance and the power mismatch level inside the islands, the quality of every single island can be assessed. As can be seen in Figure (5), no further action is carried, the system loses stability at 25.25 sec. The simulation in DIgSILENT suggests out of step at t=25.25 sec for generators. As observed, the system is separated into three unbalanced groups. The frequency of the generators and the out-of-step condition of the power network and to proceed with the separation methodology, we consider t=25.55 sec, the loss of synchronism in the system, as the point when the system separation is applied. After, obtaining the PPCC matrix and applying the separation using the Hierarchical DBSCAN, as can be seen in Figure (3), the power network is divided into three islands.

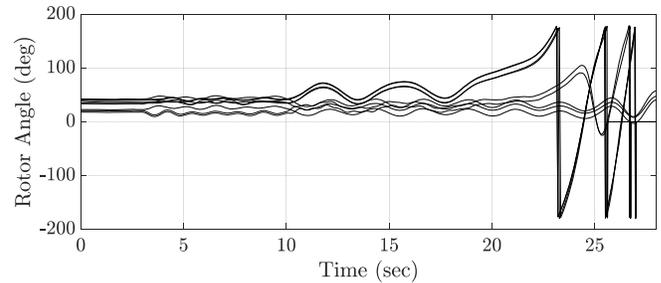

Fig. 5. Generator rotor angle oscillation during the simulation

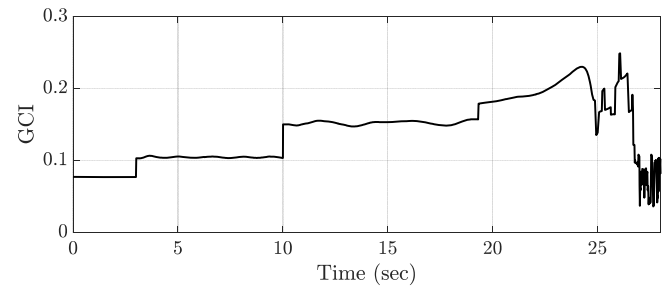

Fig. 6. Group Coherency Index (GCI) variation

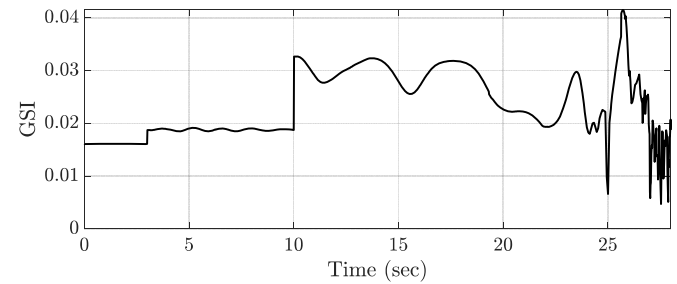

Fig. 7. Group Separation Index (GSI) variation

V. CONCLUSION

This paper proposes a novel methodology for discovering the degree of coherency among buses using the correlation index of the voltage angle at the bus to measure the strength of the association between each pair of buses and by using the Hierarchical Density-Based Spatial (HDBSCAN) clustering techniques the network was portioned to stable islands. In this methodology, the dynamic reaction of distributed generators is assessed by the phase angles variation at the buses nearby the generator. Further, the strength of the buses coherency was assessed, the power network integrity indices were discovered, and the overall system status was examined. It was evident from the results that this approach can determine the degree of coherency among any pair of buses i.e., generators and non-generator buses, accurately using the correlation coefficient among the buses. Further, by identifying various degrees of coherency for forming areas, the network partitioning to stable islands.